# CMB Maps at 0°.5 Resolution II: Unresolved Features

A. Kogut[1,2], G. Hinshaw[1], and C.L. Bennett[3]

## ABSTRACT

High-contrast peaks in the cosmic microwave background (CMB) anisotropy can appear as unresolved sources to observers. We fit simulated CMB maps generated with a cold dark matter model to a set of unresolved features at instrumental resolution 0°.5 to 1°.5 and present the integral density per steradian $n(>|T|)$ of unresolved features brighter than threshold temperature $|T|$. A typical medium-scale experiment observing 0.001 sr at 0°.5 resolution would expect to observe one unresolved feature brighter than 85 $\mu$K after convolution with the beam profile, with less than 5% probability to observe a source brighter than 150 $\mu$K. Increasing the power-law index of primordial density perturbations $n$ from 1 to 1.5 raises these temperature limits $|T|$ by a factor of 2. The results are only weakly dependent on the assumed values of the Hubble constant and baryon density.

*Subject headings:* cosmic microwave background

[1] Hughes STX Corporation, Laboratory for Astronomy and Solar Physics, Code 685, NASA/GSFC, Greenbelt MD 20771

[2] E-mail: kogut@stars.gsfc.nasa.gov

[3] Laboratory for Astronomy and Solar Physics, Code 685, NASA Goddard Space Flight Center, Greenbelt MD 20771

## 1. Introduction

Measurements of the angular distribution of the cosmic microwave background radiation are an effective probe of conditions in the early universe. On angular scales above a few degrees, the fluctuations are larger than the particle horizon at recombination and probe the unperturbed primordial density distribution. On medium angular scales (near $0°.5$), causal mechanisms affect the CMB anisotropy. The detection of CMB anisotropy at 7° resolution by the *Cosmic Background Explorer* (Smoot et al. 1992, Bennett et al. 1994) has focused attention on experiments at smaller angular scales, where the statistical power to discriminate between various models of structure formation is greater (Efstathiou, Bond, & White 1992; Coulson et al. 1994). Several groups have reported detections of anisotropy on degree angular scales: ACME (Schuster et al. 1993), ARGO (De Bernardis et al. 1994), Big Plate (Wollack et al. 1994), MAX4 (Devlin et al. 1994), MSAM (Cheng et al. 1994), and Python (Dragovan et al. 1994). Particular interest has been generated by the MSAM collaboration's detection of bright unresolved features consistent with a CMB spectrum. All recent medium-scale results suffer from limited sky coverage, typically 0.01 sr or less, which introduces uncertainties interpreting the data. Are bright unresolved sources evidence of foreground contamination, instrumental artifacts, new cosmological information, or are they an expected feature in the standard scale-invariant cold dark matter model? In this Letter, we address the question of unresolved features in the CMB, and attempt to quantify the number density per steradian of such unresolved sources as a function of intensity and cosmological model.

## 2. Analysis

We simulate a series of CMB maps at $0°.5$ resolution FWHM (full width at half-maximum), each consisting of 4096 equal-area pixels within a 10.4° × 10.4° patch on the sky. We express the CMB temperature field observed at instrumental beam dispersion $\sigma_b$ as a sum over spherical harmonics, $\Delta T(\theta,\phi) = \sum_{\ell,m} a_{\ell m} Y_{\ell m}(\theta,\phi) e^{-\ell(\ell+1)\sigma_b^2}$, and generate the amplitudes $a_{\ell m}$ for the range $\ell = [2, 500]$ as Gaussian random variables of zero mean and $\ell$-dependent variance appropriate to the model under consideration (Stompor 1994, Bond & Efstathiou 1987). Hinshaw et al. (1994) discuss the simulated maps in more detail.

Each simulated map is fitted to a series of unresolved sources using a peak normalization such that the fitted amplitude $T$ represents the peak amplitude of the unresolved feature in the sky after convolution with the instrument beam profile. The fitting algorithm searches the map for the brightest pixel, calculates the mean of the 80 pixels in a 9 × 9 array surrounding the brightest pixel, and fits the mean-subtracted patch to a Gaussian profile with the same instrumental beam dispersion ($0°.5$ FWHM) used to generate the map. The pixels are searched in order of decreasing temperature, with each fitted profile removed from the map before proceeding to the next pixel.

We add no instrument noise to the maps, but estimate the "confusion noise" as the standard deviation of the 80 pixels in each patch as a guide to the significance of each fitted Gaussian profile. We repeat the analysis for instrumental resolution $0°.5$, $1.0°$, and $1°.5$, using a larger patch ($20° \times 20°$) for the two larger beam widths. The fitted amplitudes $T$ and uncertainties $\delta T$ for all pixels in 100 realizations for each model form a data base which we use to evaluate the integral source density $n(>|T|)$ of sources per steradian brighter than threshold $|T|$.

Figure 1 shows a contour plot of a typical realization for spatially flat cold dark matter (CDM) models, with baryon density $\Omega_b = 0.05$, Hubble constant $H_0 = 50$ km s$^{-1}$ Mpc$^{-1}$ and *COBE* normalization $Q_{rms-PS} = 20$ $\mu$K and $n = 1$ or $Q_{rms-PS} = 14$ $\mu$K and $n = 1.5$ (Gorski et al. 1994). Approximately 14% of the map area is covered by unresolved features brighter than 5 times the local "confusion noise." The chopping strategies employed by medium-scale experiments act as a high-pass filter on the sky; consequently, these features would be detected as unresolved CMB sources by an experiment with sufficiently low instrumental noise. Enough of the map area lies within one beam width of an unresolved feature for their detection not to be a statistically unlikely event.

Figure 2 shows the integral number density per steradian $n(>|T|)$ of fitted amplitudes brighter than threshold $|T|$ for CDM models with $n$=1.0 and 1.5. For $n$=1, an experiment covering 0.01 sr of the sky would expect to observe on average one unresolved feature brighter than 135 $\mu$K, with less than 5% probability to observe a source brighter than 175 $\mu$K. An experiment with smaller sky coverage would observe correspondingly fewer features: at $10^{-3}$ sr, a typical experiment would observe one source brighter than 85 $\mu$K, with 5% probability to observe a source brighter than 150 $\mu$K. The source densities depend somewhat on the cosmological model: changing the index $n$ from 1 to 1.5 increases the fitted amplitudes by a factor 2.2, while changing the baryon density $\Omega_b$ from 0.05 to 0.2 increases the amplitudes by 1.5.

Table 1 lists recent results from medium-scale observations. Of these experiments, only the MSAM collaboration reports a detection of an unresolved feature consistent with a CMB spectrum. For the other experiments, column 4 ($T_{max}$) lists either the brightest unresolved source regardless of its fitted spectrum (ACME) or the brightest pixel convolved with the point-source response function of the chop strategy employed by that instrument. Selecting the brightest pixel in place of the brightest fitted feature for groups not reporting source detections is clearly only an approximation, but our simulations indicate it is reasonable. The normalization in temperature units of features smaller than the instrumental resolution depends on the beam solid angle. We adopt without change all results reported in temperature units; for results in flux units (e.g. MSAM) we normalize using the Gaussian FWHM in column 2.

These brightest sources may be compared to the integral source densities from our simulations. Column 5 lists the mean number of sources expected within the measured solid angle $\Omega$ as bright or brighter than the measured $T_{max}$, as determined for a flat CDM model ($\Omega_b = 0.05$, $h = 0.5$, $n$=1) with beam resolution matched to the specific experiment. We derive a mean value for each experiment by integrating the source curves over an appropriate range of $T_{max}$,

$$\langle N(|T|>T_{max})\rangle = \frac{\Omega \int n(>|T|)G(T)dT}{\int G(T)dT},$$

where $G(T)$ is a Gaussian centered on each $T_{max}$, reflecting the contribution of instrument noise and absolute calibration uncertainties. The quoted uncertainties reflect the sample variance deduced by shifting the curves $n(>|T|)$ by one standard deviation. Values for the expected mean number of sources significantly larger (smaller) than unity indicate that the experiment observes peak anisotropy smaller (larger) than expected. Details of the chop strategy and scan pattern are not expected to significantly modify these values. Column 6 lists the brightest expected source for each experiment, e.g., the amplitude $T$ for which the mean number of sources in the given solid angle equals unity. This can be compared to the observed brightest source $T_{max}$ (Column 4).

Three of the experiments (ACME, Big Plate, and MSAM) observe a brightest source close to the value expected for the standard $n = 1$ CDM model. ARGO observes a paucity of bright sources given the relatively large solid angle, while MAX4 and Python observe peak amplitudes brighter then expected, closer to the $n = 1.5$ model than $n = 1$ if the *COBE* normalization is retained. Similar results arise from analysis of the power spectra (e.g. Wright et al. 1994), which utilize the full data sets instead of the local "brightest pixel" test used here.

Table 1 indicates that the detection of isolated unresolved features in the CMB is not uncommon for standard cosmological models. While observers are understandably cautious about isolated bright features in their data, these features are expected in standard models and should be included in statistical analyses. More complete sky coverage at medium angular scales would improve the constraints on cosmological models by improving the sampling statistics.

We thank Radek Stompor for providing digitized transfer functions for various CDM models, and Tony Banday for many useful discussions.

Table 1: Brightest Observed and Expected Sources For Recent Medium-Scale Results

| Experiment | FWHM | Solid Angle (sr) | $T_{max}^a$ ($\mu$K) | $\langle N(|T|>T_{max})\rangle^b$ | $T_{N=1}^c$ ($\mu$K) |
|---|---|---|---|---|---|
| ACME      | 1°.5 | $1.5\times 10^{-2}$ | $53\pm 15$  | $2.1\pm 0.4$   | 55  |
| ARGO      | 0°.9 | $1.7\times 10^{-2}$ | $45\pm 10$  | $28\pm 2$      | 120 |
| Big Plate | 1°.4 | $1.2\times 10^{-2}$ | $70\pm 20$  | $1.6\pm 0.3$   | 65  |
| MAX4      | 0°.6 | $1.0\times 10^{-3}$ | $200\pm 60$ | $0.10\pm 0.01$ | 80  |
| MSAM      | 0°.5 | $2.0\times 10^{-3}$ | $118\pm 18$ | $0.6\pm 0.2$   | 100 |
| Python    | 0°.8 | $1.1\times 10^{-3}$ | $180\pm 50$ | $0.07\pm 0.02$ | 70  |

[a] Brightest unresolved source or brightest pixel convolved with point-source response function (thermodynamic temperature).

[b] Mean number of sources brighter than $T_{max}$ in solid angle $\Omega$. The calibration and noise uncertainties in $T_{max}$ are included in the mean value; the quoted uncertainties reflect the sample variance derived from the width of the source curves.

[c] Brightest expected source for each experiment (value $|T|$ for which the mean expected source count is unity).

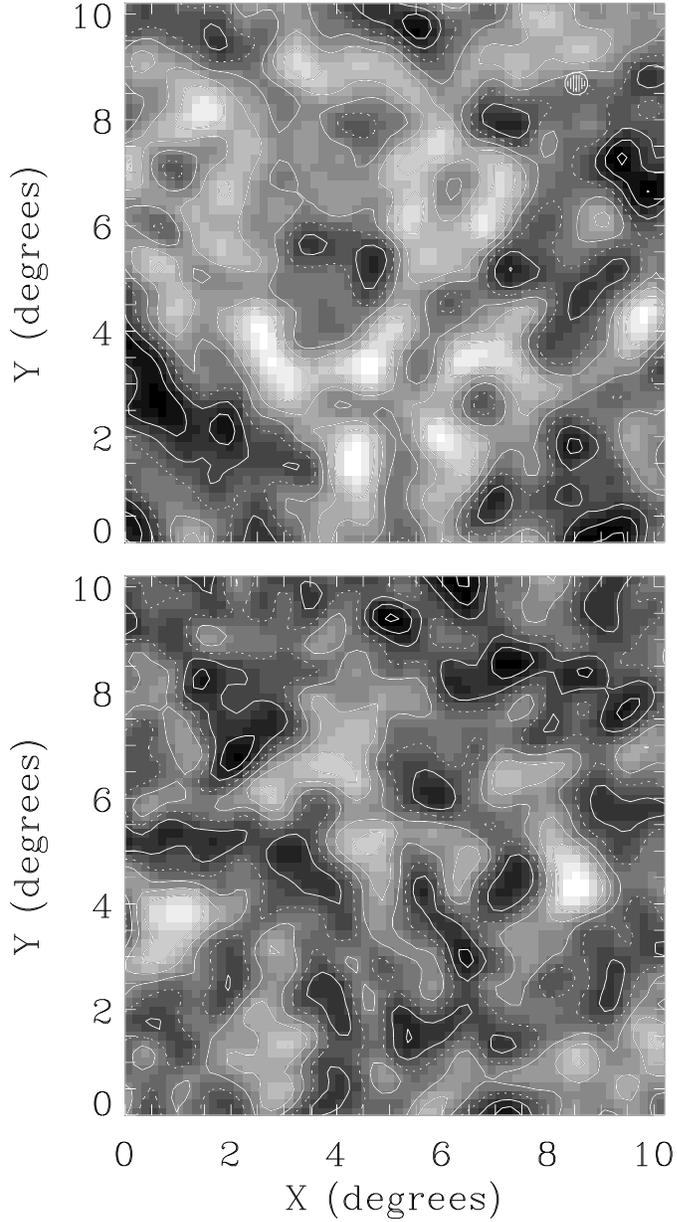

Figure 1: Contour maps of simulated CMB anisotropy in a 10° × 10° patch of the sky for CDM models with $h = 0.5$ and $\Omega_b$=0.05. The hatched disk at the upper right indicates the 0°.5 beam width. Approximately 14% of the map area is within 0°.5 of a bright unresolved feature. (top) $n$=1 and $Q_{rms-PS} = 20$ $\mu$K. Contours are -150$\mu$K to +150 $\mu$K in steps of 50 $\mu$K. The dashed line is the zero contour. (bottom) $n$=1.5 and $Q_{rms-PS} = 14$ $\mu$K. Contours are -300$\mu$K to +300 $\mu$K in steps of 100 $\mu$K.

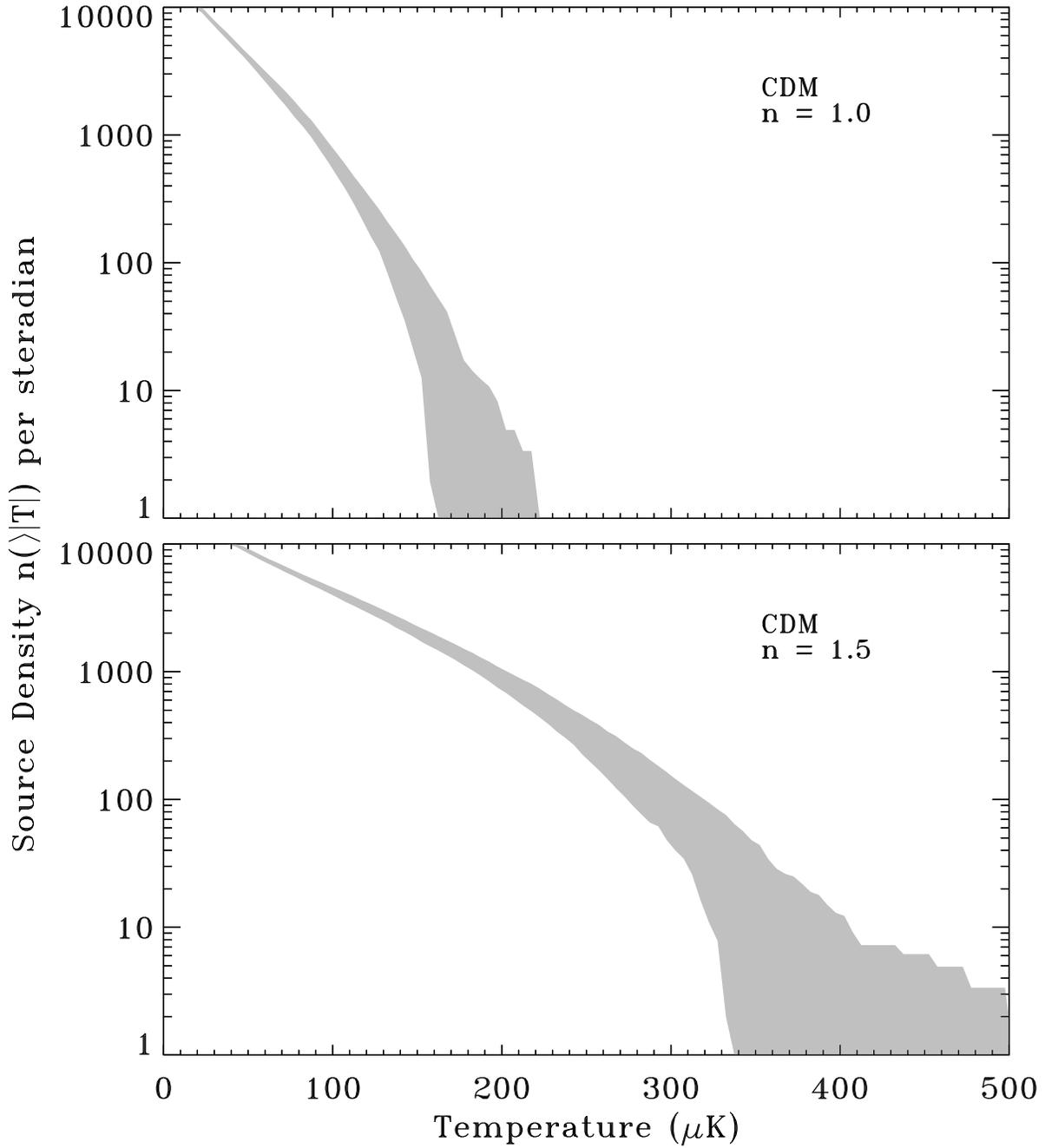

Figure 2: Integral source density per steradian $n(>|T|)$ for fitted source amplitudes brighter than threshold $|T|$ observed at instrumental resolution $0°\!.5$ FWHM. The gray band is the 68% confidence interval from 100 noiseless realizations. (top) CDM model with $h=0.5$, $\Omega_b = 0.05$, $Q_{rms-PS} = 20$ $\mu$K and $n = 1.0$ (bottom) CDM model with $h=0.5$, $\Omega_b = 0.05$, $Q_{rms-PS} = 14$ $\mu$K and $n = 1.5$. The normalization $Q_{rms-PS}$ is chosen to match the $COBE$ data at each value of $n$.
8